\documentclass[twocolumn,preprintnumbers,superscriptaddress,amsmath,amssymb]{revtex4-2}
\usepackage{amsfonts}
\usepackage{amsmath}
\usepackage{array}
\usepackage{dcolumn}
\usepackage{epsfig}
\usepackage{graphicx}
\usepackage{subfigure}
\usepackage{dcolumn}
\usepackage{bm}
\usepackage{times}
\usepackage{xcolor}
\usepackage{float}
\usepackage{soul}
\usepackage{comment}
\usepackage{enumerate}
\usepackage{multirow}
\usepackage{xr}
\usepackage{mathrsfs}
\usepackage[normalem]{ulem}
\usepackage{scalerel}

\newcommand{\CellWithForcedBreak}[2][c]{\begin{tabular}[#1]{@{}c@{}}#2\end{tabular}}

\newcolumntype{C}[1]{>{\centering\let\newline\\\arraybackslash\hspace{0pt}}m{#1}}

\begin{document}

\title{Spreading dynamics of information on online social networks}
\author{Fanhui Meng}
\affiliation{School of Systems Science and Engineering, Sun Yat-sen University, Guangzhou, 510275, China}
\author{Jiarong Xie}
\affiliation{Center for Computational Communication Research, Beijing Normal University, Zhuhai, 519087, China}
\affiliation{School of Journalism and Communication, Beijing Normal University, Beijing, 100875, China}
\affiliation{School of Computer Science and Engineering, Sun Yat-sen University, Guangzhou, 510006, China}
\author{Jiachen Sun}
\affiliation{School of Computer Science and Engineering, Sun Yat-sen University, Guangzhou, 510006, China}
\author{Cong Xu}
\affiliation{Department of Statistics and Data Science, College of Science, Southern University of Science and Technology, Shenzhen, 518055, China}
\author{Yutian Zeng}
\affiliation{Department of Statistics and Data Science, College of Science, Southern University of Science and Technology, Shenzhen, 518055, China}
\author{Xiangrong Wang}
\affiliation{College of Mechatronics and Control Engineering, Shenzhen University, Shenzhen, 518060, China}
\author{Tao Jia}
\affiliation{College of Computer and Information Science, Chongqing Normal University, Chongqing, 401331, China}
\affiliation{College of Computer and Information Science, Southwest University, Chongqing, 400715, China}
\author{Shuhong Huang}
\affiliation{Institute of Neuroscience, Technical University of Munich, Munich, 80802, Germany}
\author{Youjin Deng}
\affiliation{Department of Modern Physics, University of Science and Technology of China, Hefei, 230026, China}
\affiliation{MinJiang Collaborative Center for Theoretical Physics, College of Physics and Electronic Information Engineering, Minjiang University, Fuzhou, 350108, China}
\author{Yanqing Hu}
\email{huyq@sustech.edu.cn}
\affiliation{Department of Statistics and Data Science, College of Science, Southern University of Science and Technology, Shenzhen, 518055, China}
\affiliation{Center for Complex Flows and Soft Matter Research, Southern University of Science and Technology, Shenzhen, 518055, China}

\date{December 5, 2024}

\begin{abstract}
Social media is profoundly changing our society with its unprecedented spreading power. Due to the complexity of human behaviors and the diversity of massive messages, the information spreading dynamics are complicated, and the reported mechanisms are different and even controversial. Based on data from mainstream social media platforms, including WeChat, Weibo, and Twitter, cumulatively encompassing a total of 7.45 billion users, we uncover a ubiquitous mechanism that the information spreading dynamics are basically driven by the interplay of social reinforcement and social weakening effects. Accordingly, we propose a concise equation, which, surprisingly, can well describe all the empirical large-scale spreading trajectories. Our theory resolves a number of controversial claims and satisfactorily explains many phenomena previously observed. It also reveals that the highly clustered nature of social networks can lead to rapid and high-frequency information bursts with relatively small coverage per burst. This vital feature enables social media to have a high capacity and diversity for information dissemination, beneficial for its ecological development.
\end{abstract}

\maketitle
\section{Significance}
We find that in the information spreading process of online social networks, the social reinforcement effect and the social weakening effect coexist, and this is a universal phenomenon. We propose a concise mathematical model which can well describe the empirical spreading dynamics. Our theory indicates that the highly clustered nature of the social network structure results in high-frequency information bursts with relatively small coverage, enabling social media to have high capacity and diversity for information dissemination.

\section{Introduction}
With the advantages of open, easily accessible, and real-time communication social media has given rise to new social communities involving billions of people across borders, races, and cultures~\cite{wang2009understanding, lorenz2019accelerating, lazer2009computational}. It plays a pivotal role in almost all aspects of human societies~\cite{rajkumar2022causal, wang2022weak, chetty2022social1, lehmann2018complex, deville2016scaling}, including public health~\cite{gallotti2020assessing}, political campaigns~\cite{bovet2019influence}, e-commerce~\cite{kempe2003maximizing, kitsak2010identification, hu2018local, morone2015influence}, awareness of societal issues~\cite{chetty2022social1, bakshy2015exposure}, information security~\cite{bovet2019influence, del2016spreading}, and so on. Social media has profoundly changed the way people share information. Hence, understanding the information spreading mechanisms, and especially how global collective spreading behaviors are triggered by individuals, has attracted extensive attention in recent years~\cite{watts2002simple, kwak2010twitter, bakshy2012role, del2016spreading, vosoughi2018spread, hebert2020macroscopic, gleeson2014competition}.

In earlier studies~\cite{castellano2009statistical, vespignani2012modelling, iribarren2009impact, goel2016structural, gruhl2004information, liu2015events, moreno2004dynamics}, the dynamics of information spreading were believed to be analogous to the spread of disease in a population. Later, with evidence from empirical experiments~\cite{centola2010spread, bakshy2012role}, it was realized that this is not the case. Specifically, in 2010, Centola observed in an online propagation experiment that the more times people see a piece of information, the more willing they are to spread it~\cite{centola2010spread}. Accordingly, the social reinforcement effect in information spreading on online social networks was proposed. Bakshy et al. repeatedly observed the reinforcement effect through online Facebook experiments~\cite{bakshy2012role}. Prior to online social media, this interesting phenomenon was observed earlier in offline experiments dating back to Granovetter~\cite{granovetter1978threshold} and has been intriguing over the past decade~\cite{vespignani2012modelling, lehmann2018complex, myers2012clash, hebert2020macroscopic, chetty2022social1}.

However, our empirical observations of data, generated naturally in the real world instead of human-designed experiments, show that the spreading dynamics exhibit complex patterns~(see Fig.~\ref{Fig1}A), far beyond the simple scenario of the social reinforcement effect. For example, as the number of exposures increases, the retweeting probability for some messages first increases and then saturates while that for some other messages increases rapidly followed by a dramatic decrease~(see Fig.~\ref{Fig1}A). Actually, reliable observations of real-world natural spreading reported by other researchers also demonstrate controversies with the social reinforcement effect~\cite{hebert2020macroscopic, davis2020phase, hebert2015complex, romero2011differences, myers2012clash, ugander2012structural, cheng2018diffusion}. Romero et al. found that the spread of information on Twitter is not entirely the result of social reinforcement as multiple exposures would reduce the spreading probability~\cite{romero2011differences}; Ugander et al. observed on Facebook that the higher the proportion of common neighbors between users, the significantly lower the probability of transmission~\cite{ugander2012structural}; Cheng et al. also found that the likelihood of forwarding a message typically decreases with the number of exposures on Facebook~\cite{cheng2018diffusion}. 

Moreover, an evident paradox arises between the theory of the reinforcement effect and the empirical propagation of information on large-scale social networks. On the one hand, the reinforcement reflects a positive mutual feedback between the number of exposures and the forwarding probability. As information spreads and the number of times users receive it increases~\cite{centola2010spread}, the retweet probabilities of potential spreaders are enhanced, and, as feedback, the number of exposures for non-spreading users is further increased. This would commonly lead to a flood of information, with nearly all users retweeting the same information. On the other hand, the total number of messages forwarded daily is limited under the reinforcement effect as it would require enormous retweets per day per user to stimulate numerous messages on a large scale, while the time users spend on social media is physiologically limited. However, the fact is that tons of messages are spreading on social networks every day, but barely any on an overwhelmingly large scale~(even outbreak sizes are tiny compared to the size of the entire network, e.g., at most $0.04\%$, see Table~\ref{table:dataset}).

Here, we study the spreading dynamics of online social networks from a new perspective, recognizing that the propagation process of a piece of information on social media is essentially the spreading process of people's behavior related to the information. We illustrate this perspective by comparing information spreading with the spread of the virus in a population. In virus transmission, the virus takes people as the host and people play a passive role during virus spreading, with little awareness until symptoms appear. In contrast, people have always been proactive in the process of information propagation. This can be understood as people repeating or mimicking the retweeting behavior of his/her peers. In this sense, the propagation of information can be regarded as the propagation of human behaviors. It is indicated that when studying the mechanisms of information spreading, the natural focus should mainly lie on human behavior, not only the information itself. Therefore, the spreading mechanism, in principle, deeply relies on the gains of mimicking peer behaviors via social interactions among users.

We systematically analyze the spreading trajectories of real information and the corresponding individual retweeting behaviors on three large-scale social media platforms~(see Table~\ref{table:dataset}), WeChat, Weibo, and Twitter.  A brief description of the empirical data used is provided in the Methods section and please refer to SI Sec.~I for more detailed information. We find that users' forwarding behavior on a piece of information is jointly determined by the following three factors: the intrinsic importance of the information, the average proportion of common neighbors between a user and her friends, and the users' uncertainty, that is, the users' sensitivity to the number of exposures and deviation in estimating of the proportion of common neighbors. The first indicates the inherent spreading power of the information, and the latter two represent interactions between users. 
Based on them, we formulate a concise equation for the propagation dynamics:
\begin{equation}\label{eq:unified_equation_ultimate}
  \beta_i(x) = \alpha_i x (1 - \gamma) ^ {x ^ {\omega_i}},
\end{equation}
where $\beta_i(x)$ is the retweeting probability of message $i$ at the $x$th exposure to a user, $\alpha_i$, $\gamma$, $\omega_i$ respectively describe the three factors mentioned above.

Note that although there are three parameters in Eq.~(\ref{eq:unified_equation_ultimate}), $\omega_i$ is the only one that has to be fitted while $\alpha_i$ and $\gamma$ can be directly measured from the empirical spreading trajectories and network structure, respectively. We find that this equation well captures most of the spreading trajectories of empirical information~(see Fig.~\ref{Fig1}B-D). It also resolves a number of controversial claims, including the contradiction of the social reinforcement mechanism, social weakening effects~\cite{romero2011differences}, weak ties promoting spreading~\cite{bakshy2012role}, and more common neighbors reducing the spreading transmissibility~\cite{ugander2012structural}, among other unexplained phenomena by previous theories. Furthermore, the equation suggests that the highly clustered nature of social media leads to high-frequency information bursts and a small coverage per burst. This endows social media with high throughput and diversity for information dissemination, which is super beneficial to its ecological sustainability.

\section{Results}
\subsection{Retweeting probability}
\begin{figure*}[!htbp]
  \centering
  \includegraphics[width=17cm]{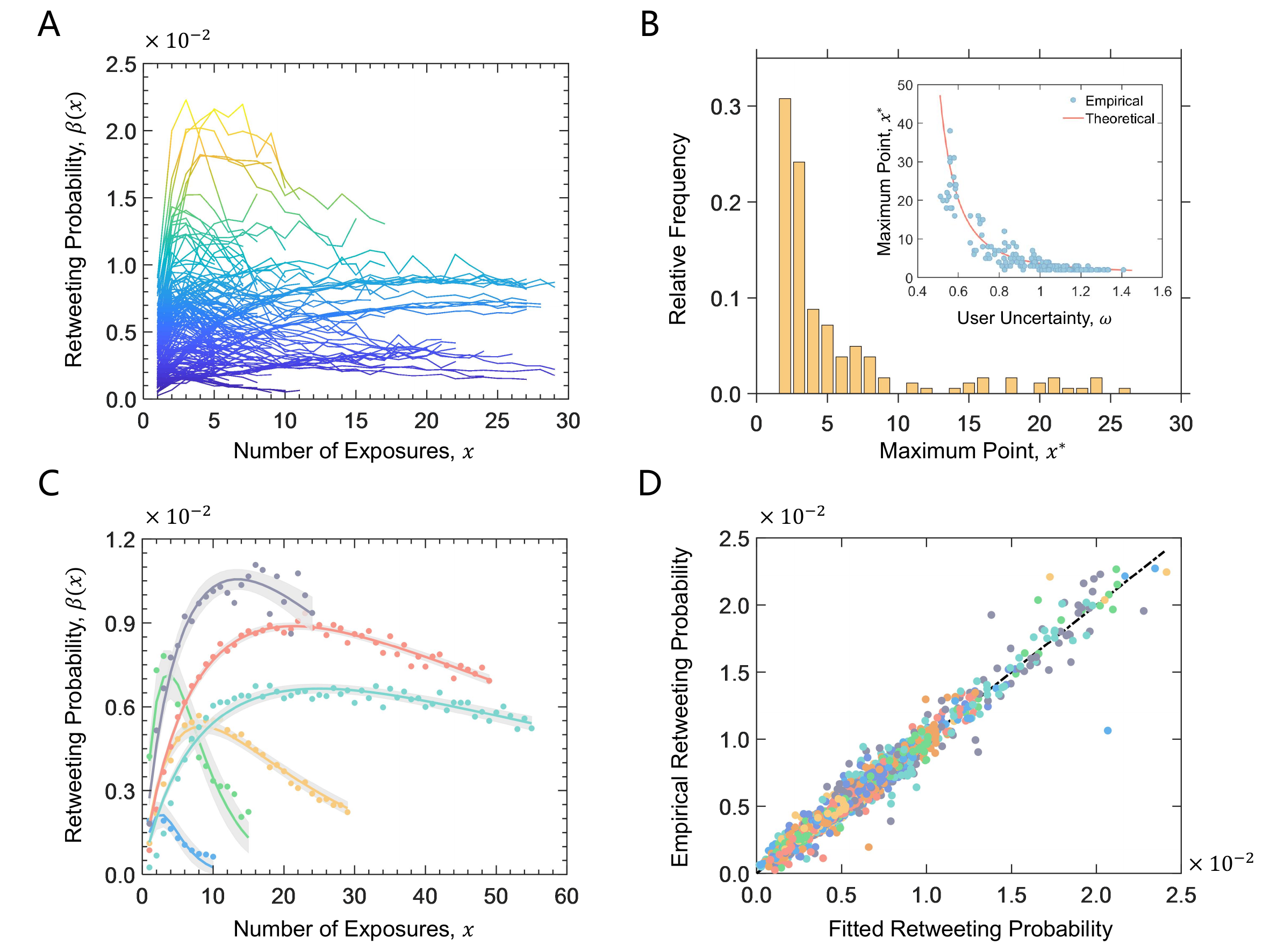}
  \caption{
  \textbf{The empirical and fitted retweeting probabilities.}
  \textbf{(A)} The empirical spreading trajectories of messages collected from WeChat. Each solid line represents the retweeting probability $\beta(x)$ versus the number of exposures $x$ for a single message.
  \textbf{(B)} The distribution of $x^*$, i.e., the number of exposures $x$ at which $\beta(x)$ reaches its maximum, based on the WeChat messages. Typically, $\beta(x)$ is maximized at $x^*$ = 2 or 3, in total accounting for 55\% of all cases. Inset: the agreement between theoretical and empirical results of $x^*$~(see SI Sec.~III).
  \textbf{(C)} The empirical~(dots) and fitted~(solid lines) retweeting probabilities of 6 representative messages from WeChat by fitting Eq.~(1) on the empirical spreading trajectories. The gray shaded area is the 95\% confidence interval. Please see SI Sec.~IV for the rest of fitted Weibo trajectories. 
  \textbf{(D)} The empirical versus fitted retweeting probabilities for the WeChat messages. Dots of the same color represent data of the same message. All dots align close to the diagonal line, suggesting a quite good fit to the empirical data by Eq.~(1).
  }\label{Fig1}
\end{figure*}

The process of information spreading on social media is driven by the behavior of users: they tend to forward a message to their friends after seeing it multiple times~\cite{centola2010spread}. Thus, the interaction associated with retweeting behaviors is the core of the propagation dynamics. Formally, we define $\beta_i(x)$ of message $i$ as the proportion of users who perform the forwarding operation at the $x$th exposure, which is expressed as
\begin{equation}\label{eq:beta_x}
  \beta_i(x) = \frac{m_i(x)}{r_i(x)},
\end{equation}
where $m_i(x)$ is the number of users who forward the information at the $x$th exposure, and $r_i(x)$ is the number of users who have been exposed to the information no less than $x$ times and have not yet forwarded it at the first $x-1$ times~(see the Methods section and SI Sec.~II). For messages spread on a relatively large scale, we are able to accurately measure the retweeting probability $\beta(x)$, a generic notation of $\beta_i(x)$. Our empirical data clearly demonstrate diverse and complicated spreading patterns~(see Fig.~\ref{Fig1}A).

\begin{table*}[!htbp]\centering
  \caption{
  \textbf{Description of the major datasets used.} ``Information outbreak size'' is the range of the number of users retweeting each message; ``Total number of exposed users'' is the sum of the number of users exposed to each message; ``Total number of retweeting users'' is the sum of the number of users retweeting each message. Please refer to SI Sec.~I for more detailed data descriptions and SI Sec.~V for an explanation of how exposures are measured. Here, all the information we study is publicly accessible.
  }\label{table:dataset}
  \begin{tabular}{ m{12mm} C{28mm} C{22mm} C{28mm} C{22mm} C{28mm} C{28mm}} 
  \toprule
  Platforms & Time period & Spreading scale & \CellWithForcedBreak{Information \\ outbreak sizes} & \CellWithForcedBreak{Number of \\ messages}{} & \CellWithForcedBreak{Total number of \\ exposed users} & \CellWithForcedBreak{Total number of \\ retweeting users} \\ \hline
  WeChat & 2019.06 - 2019.10 & Large-scale {} & 10,733 - 468,970 & 182 & 5,412,813,032 & 23,983,694 \\ \hline
  \multirow{2}{*}{Weibo} & \multirow{2}{*}{2017.01 - 2017.10} & Large-scale & 3,235 - 35,045 & 100 & 2,033,741,843 & 1,189,138 \\
                                                          &   & Small-scale & 1 - 100 & 7,946 & 4,346,107,108 & 91,627 \\ \hline
  Twitter & 2010.10 & Small-scale & 1 - 100 & 11,072 & 16,945,294 & 81,633 \\ \hline \hline
  \end{tabular}{}
\end{table*}

\subsection{Unified equation for propagation dynamics}
\begin{figure*}[!htbp]
  \centering
  \includegraphics[width=15cm]{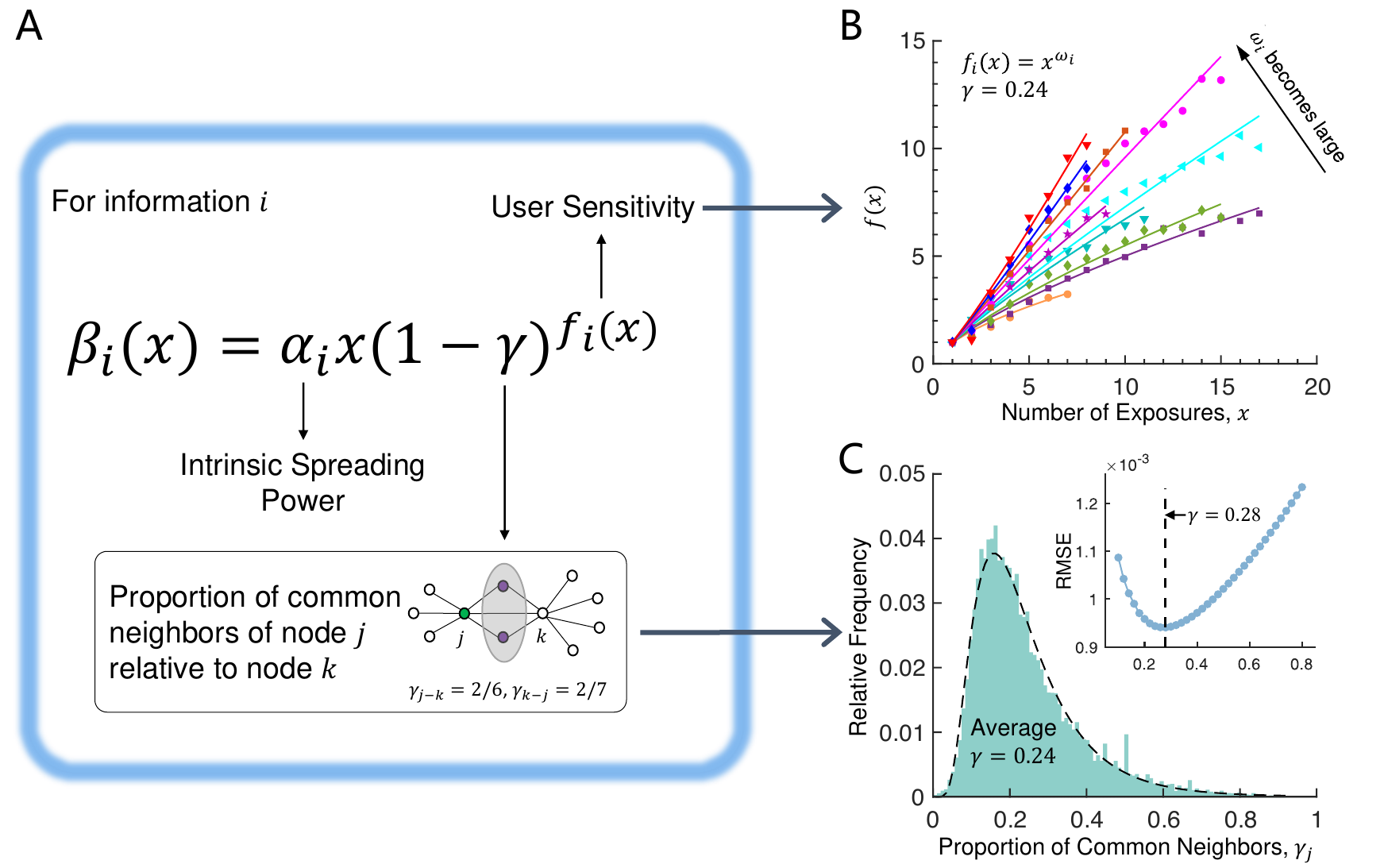}
  \caption{
  \textbf{The unified spreading equation.}
  \textbf{(A)} The unified spreading equation and the interpretation of each term for a single message $i$. Parameters $\alpha_i$, $\gamma$ and the term $f_i(x)$ correspond to the intrinsic spreading power, the average proportion of common friends between a user and its neighbors~(see Sec.~XIV B), and the users' uncertainty to the gains of retweeting message $i$, respectively. To be more specific, $\alpha_i x$ describes a linear social reinforcement effect; $(1-\gamma)^{f_i(x)}$ formulates the calibration of the effective exposure rate, i.e., the proportion of a user's friends who would receive a message for the first time if the user retweets the message.
  \textbf{(B)} The explicit form of $f_i(x)$ as $x^{\omega_i}$ obtained from empirical WeChat data. Dots of the same color represent data of the same message. The solid line shows a theoretical fitting of $f_i(x)=x^{\omega_i}$ obtained by Eq.~(\ref{eq:f_i_x}). Note that while the solid lines seem to display a linear relationship between $f_i(x)$ and $x$, i.e., $f_i(x)=\omega_i x$, this is actually not the case as the boundary condition $f_i(1)=1$ is not satisfied under $f_i(x)=\omega_i x$.
  \textbf{(C)} The distribution of $\gamma$ measured directly from the network structure of WeChat, which follows a log-normal distribution with mean $0.24$. Inset: the estimated $\gamma$ by fitting to empirical WeChat trajectories. The calculation of $\gamma$ is illustrated by an example in~(A), the proportion of common friends of node $j$ with its neighbor $k$ is $\gamma_{j-k}=2/6$, averaging over all neighbors of $j$ yields $\gamma_j$, then averaging over all node $j$ finally gives $\gamma$.
  }\label{Fig2}
\end{figure*}

To quantitatively describe the complicated pattern of $\beta(x)$, we begin with a more general form of Eq.~(\ref{eq:unified_equation_ultimate}), by defining $\beta_i(x) = \alpha_i x (1 - \gamma) ^ {f_i(x)}$~(see Fig.~\ref{Fig2}). In this equation, $\alpha_i$ represents the intrinsic importance of message $i$; the second multiplier $x$ describes the basic linear reinforcement effect; $\gamma$ is the average proportion of common neighbors between any two users; $f_i(x)$ is a power index for message $i$ used to calibrate the effective exposure rate as the result of users' retweeting, which can be understood as users' uncertainty to the gains of retweeting the message. The specific case of Eq.~(\ref{eq:unified_equation_ultimate}) in which $f_i(x)$ is determined to be $x^{\omega_i}$ would be discussed later.

Eq.~(\ref{eq:unified_equation_ultimate}) encompasses the three core elements involved in information spreading, which form two competing mechanisms. On the one hand, when a user receives the same information multiple times, the perceived importance of the information increases, leading to the social reinforcement mechanism described by $\alpha_i x$. On the other hand, when a user receives the same information multiple times~(typically from different friends), on average many of her friends should have already been exposed to the information. This would reduce the proportion of potential ``fresh audiences''~(who receive the information for the first time) if the user forwards the information, as described by $(1-\gamma)^{f_i(x)}$. In summary, the information spreading mechanism represented by $\beta(x)$ is formulated from the perspective of human spreading behaviors, i.e., $\beta(x)$ is the gain~(which equals the product of information importance and proportion of potential audiences) that motivates users' forwarding behavior.

For better understanding, we exemplify the above form of $\beta_i(x)$ via three simple and hypothetical cases. First, consider the case when $\gamma=1$, i.e., every pair of users is connected in a network. If a user posts a message, then all users would be exposed to the message, making it valueless~(zero gain) for another user to retweet the message. This is consistent with the result of Eq.~(\ref{eq:unified_equation_ultimate}), i.e., $\beta_i(x) \equiv 0$. Second, for the case when $f_i(x) \equiv 1$, Eq.~(\ref{eq:unified_equation_ultimate}) reduces to a simple linear reinforcement effect. However, $f_i(x)$ should depend on $x$ in general as the more times a message appears in one's social network, the more of her friends are expected to be aware of the message already. Third, assuming that in a social network, every two users share $\gamma$ proportion of common neighbors and each user posts messages independently, then only $(1-\gamma)^x$ proportion of a user's friends would remain unaware of a message if it is posted by $x$ of the user's friends. This corresponds to the case when $f_i(x)=x$ in Eq.~(\ref{eq:unified_equation_ultimate}). As will be shown later, the form of $f_i(x)$ may not be perfectly linear but can be super-linear or sub-linear in $x$ under cases in reality.

\subsection{Form of the power index}
The form of the power index $f_i(x)$ can be nicely determined by combining theoretical and empirical analyses. From the theoretical perspective, we use the boundary condition that $f_i(1)=1$. This comes from the fact that when a user is exposed to a message for the first time~($x=1$), $\gamma$ proportion of her neighborhood that overlaps with the message poster is also aware of the message, yielding an effective exposure rate of $1-\gamma$ if the user forwards the message. A direct result of the boundary condition is $\beta_i(1)=\alpha_i (1-\gamma)$. Then for $x>1$, we plug $\alpha_i=\beta_i(1) / (1-\gamma)$ into Eq.~(\ref{eq:unified_equation_ultimate}) and obtain
\begin{equation}\label{eq:f_i_x}
  f_i(x) = \frac{\ln \left[ \beta_i(x) / \beta_i(1) \right] - \ln x}{\ln (1-\gamma)} + 1.
\end{equation}

Eq.~(\ref{eq:f_i_x}) enables us to study the properties of $f_i(x)$ via empirical spreading trajectories. Specifically, from the empirical spreading trajectory of message $i$, we obtain the observed retweeting probability $\beta_i(x)$ via Eq.~(\ref{eq:beta_x}) and plug it into Eq.~(\ref{eq:f_i_x}) to get the empirical value of $f_i(x)$~(see Fig.~\ref{Fig2}B). Note that the value of $\gamma$ used in the computation above is $\gamma=0.24 \pm 0.001$, obtained as the average proportion of common neighbors between any two users based on the network structure of WeChat, as shown by Fig.~\ref{Fig2}C. It turns out that $f_i(x)$ can be well fitted by $f_i(x)=x^{\omega_i}$ while other straightforward functions, such as linear functions, cannot adequately capture its variability~(see discussions in SI Sec.~VI). Consequently, the equation for propagation dynamics is eventually established as Eq.~(\ref{eq:unified_equation_ultimate}). Actually, the classical SIR model can be recovered from Eq.~(\ref{eq:unified_equation_ultimate}) under the scenario of random network at small spreading coverage~(see SI Sec.~VII A).

\begin{figure*}[!htbp]
  \centering
  \includegraphics[width=17cm]{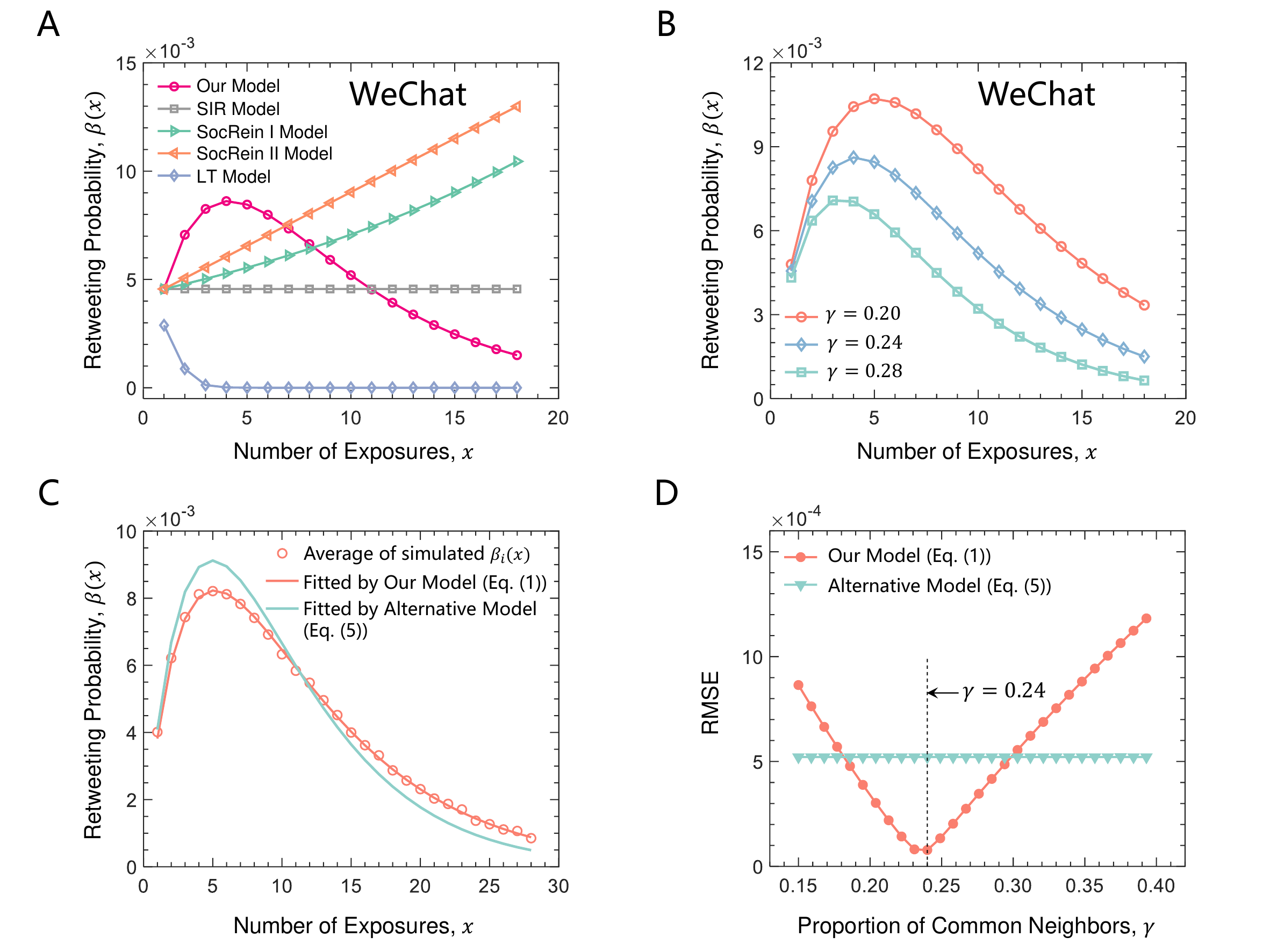}
  \caption{
  \textbf{Theoretical retweeting probability curves, weak tie effect and model comparison.}
  \textbf{(A)} The theoretical retweeting probability curve under different models. The SIR model assumes a constant retweeting probability, i.e., $\beta(x)\equiv\beta$. ``SocRein I model'' and ``SocRein II model'' refer to two social reinforcement models with retweeting probabilities $\beta(x)=\beta(1) (1+b)^{x-1}$~\cite{chen2020rumor} and $\beta(x)=1-(1-\beta(1)) (1-b)^{x-1}$~\cite{zheng2013spreading}~($b$ is a parameter for the strength of reinforcement), respectively. Parameter settings: Our model, $\alpha=0.006$, $\gamma=0.24$, $\omega=0.95$; SIR model, $\beta=\alpha(1-\gamma)=0.00456$; SocRein I model, $b=0.05$; SocRein II model, $b=0.0005$; LT~(linear threshold) model, activation threshold $=0.025$. These parameters are set close to the typical values from the WeChat dataset~(see SI Sec.~X), and they do not affect the shape of the curves. The figure shows that only our model exhibits non-monotonicity which best describes the empirical data. 
  \textbf{(B)} The theoretical retweeting probability curve under our model with different values of $\gamma$. The value of $\gamma$ can be considered as a measure of how closely a user is connected to her friends. From this perspective, the figure shows the weak tie effect~(smaller $\gamma$) under our model.
  \textbf{(C)} Fitting average simulated retweeting probability with our model and an alternative model as Eq.~(\ref{eq:M2}). Average retweeting probability is generated from Eq.~(\ref{eq:M2}), where $\tilde{\omega}_{i}$ follows a powerlaw distribution. For other distribution assumptions, such as $\tilde{\omega}_{i}$ follows a normal and exponential distribution, consistent results can be found in SI Sec.~VI. 
  \textbf{(D)} Performance of fitting average simulated retweeting probability with our model and an alternative model. The RMSE of the alternative model has the same value for all $\gamma$ indicating a redundancy between $\gamma$ and $\omega$. The alternative model cannot automatically select the best estimate for $\gamma$, in contrast, our model demonstrates a higher level of performance, as the best-estimated value of $\gamma$ closely matches $\gamma=0.24$, which is the ground truth of the parameter $\gamma$.
  }\label{Fig3}
\end{figure*}

The superiority of our model is further demonstrated by Fig.~\ref{Fig3}A. Comparing the theoretical retweeting probability curve under different models, our model uniquely achieves a curve that first rises and then falls, which is a prevalent pattern of information spreading on social media platforms. Furthermore, our model supports the weak tie theory, that is, weaker social ties facilitate communication and information exchange. Fig.~\ref{Fig3}B shows that the retweeting probability of users with lower $\gamma$~(i.e., those who connect less closely to their friends) is significantly higher than those with higher $\gamma$. The weak tie effect cannot be explained by either the SIR model or the social reinforcement model.

\subsection{Fitting empirical data}
Eq.~(\ref{eq:unified_equation_ultimate}) with $\gamma=0.24$ well fits the retweeting probabilities of large-scale spreading WeChat messages~(see Fig.~\ref{Fig1}C-D, Fig.~S4). Fitting results for each large-scale trajectories on WeChat and Weibo are shown in Fig.~S25, S26. Details of evaluation metrics of model fitting performance can be found in SI Sec.~VIII). Other than fitting Eq.~(\ref{eq:unified_equation_ultimate}) with a fixed $\gamma$ obtained from the network structure, we can also fit the equation by treating $\gamma$ as an unknown parameter~(see details in SI Sec.~IX). Specifically, for any given value of $\gamma$, we could find the optimal $\omega_i$ and $\alpha_i$ meanwhile by fitting Eq.~(\ref{eq:unified_equation_ultimate}) to the observed spreading trajectory of message $i$ and compute the corresponding root mean square error~(RMSE). Then the value of $\gamma$ such that the average RMSE overall messages reaches its minimum is the estimated value of $\gamma$. As shown in the inset of Fig.~\ref{Fig2}C, $\gamma$ is estimated to be $0.28 \pm 0.02$~(the standard deviation is obtained by the Bootstrap method~\cite{efron1994introduction}).

Note that from a purely model-fitting perspective, $\gamma$ is just a parameter to be fitted and its physical meaning is obscure, i.e., there is no explicit reason why $\gamma$ should reflect the structure of the underlying social network for message spreading and more specifically, represent the proportion of common neighbors. Therefore, it is really surprising to find that the value of $\gamma=0.28\pm 0.02$, as estimated by fitting Eq.~(\ref{eq:unified_equation_ultimate}) to the empirical trajectories, is consistent with $\gamma=0.24\pm 0.001$, which is directly measured from the social network structure without referring to Eq.~(\ref{eq:unified_equation_ultimate}). This striking consistency provides strong support for the form of Eq.~(\ref{eq:unified_equation_ultimate}).

Under Eq.~(\ref{eq:unified_equation_ultimate}), the retweeting probability could incorporate a non-monotonic shape. Specifically, $\beta_i(x)$ first increases with $x$ because of the term $\alpha_i x$, demonstrating a social reinforcement effect. Then, as the term $(1-\gamma)^{x^{\omega_i}}$ starts to dominate, $\beta_i(x)$ turns to decrease with $x$ after reaching some turning point. The turning point can be mathematically determined as~(see SI Sec.~III for the details)
\begin{equation}\label{eq:x_star}
  x_i^* = \left(- \frac{1}{\omega_i \ln (1-\gamma)} \right) ^ {\frac{1}{\omega_i}}.
\end{equation}
The theoretical relationship between $x_i^*$ and $\omega_i$ given by Eq.~(\ref{eq:x_star})~(with $\gamma=0.24$) is displayed in the inset of Fig.~\ref{Fig1}B, showing that $x_i^*$ decreases rapidly as $\omega_i$ increases. The scatter plot of $x_i^*$ versus $\omega_i$ obtained from the empirical spreading trajectories of all WeChat messages is added to be compared with the theoretical curve. The high degree of agreement between the two again demonstrates that Eq.~(\ref{eq:unified_equation_ultimate}) fits the empirical data very well. We also plot the histogram of the empirical $x_i^*$ (see Fig.~\ref{Fig1}B) and find that $x_i^*$ is mostly 2 or 3 for the WeChat messages, i.e., users typically have the maximum retweeting probability when they are exposed to a message for the second or the third time.

\subsection{Interpretation and universality of Eq.~(\ref{eq:unified_equation_ultimate})}
Here, the interpretation and universality of the proposed Eq.~(\ref{eq:unified_equation_ultimate}) for information spreading are addressed. While $\alpha_i$ and $\gamma$ are clearly interpreted as the intrinsic spreading power of message $i$ and the average proportion of common neighbors between two users, the interpretation of $\omega_i$ is not so straightforward. To explain $\omega_i$, consider three alternative forms of the damping term in the equation, namely $(1-\gamma)^{\tilde{\omega}_{i} x}$, $(1-\tilde{\omega}_{i} \gamma)^{x}$, and $[(1-\gamma) \tilde{\omega}_{i})]^{x}$. These three forms are shown to be mathematically equivalent~(see SI Sec.~VI B) so that only the first one is discussed:
\begin{equation}\label{eq:M2}
  \beta_{i}(x) = \alpha_{i} x (1-\gamma)^{\tilde{\omega}_{i} x},
\end{equation}
where $\tilde{\omega}_{i}$ can be explained as the overall uncertainty of users to the estimation of $\gamma$ for a given message $i$. In reality, the uncertainty of different users is generally not the same~\cite{xie2021detecting,myers2014bursty,kwak2010twitter}, so let $\tilde{\omega}_{ij}$ denote the uncertainty of user $j$ for a given message $i$. For given $i$, when $\tilde{\omega}_{ij}$ follows a certain distribution, such as a power law distribution, a normal distribution, or an exponential distribution, then the average of $(1-\gamma)^{\tilde{\omega}_{ij} x}$ over all users can be well approximated by a stretched exponential function~\cite{Stretched,lindsey1980detailed,berberan2005mathematical}, which is $(1-\gamma)^{x^{\omega_{i}}}$ in our Eq.~(\ref{eq:unified_equation_ultimate})~(see {Fig.~\ref{Fig3}C-D} and SI Sec.~VI for details). This serves as an explanation of why Eq.~(\ref{eq:unified_equation_ultimate}) best fits the empirical spreading trajectories. Moreover, it follows that $\omega_i$ can be interpreted as the overall users' uncertainty about the average proportion of common neighbors for a given message $i$.

In addition to large-scale spreading messages, the performance of Eq.~(\ref{eq:unified_equation_ultimate}) on small-scale spreading messages is also investigated. A natural challenge arises for small-scale spreading messages: the scale of $\beta(x)$ is typically small, e.g., $10^{-3}$ from WeChat, so a large number of retweeters and exposures are required for stable calculation of $\beta(x)$, especially for larger $x$; however, the number of retweeters and exposures are generally not enough for these messages, resulting in significant fluctuations in the calculation of $\beta(x)$. To overcome this challenge, we propose to merge the trajectories of small-scale spreading messages by taking the average of $\beta(x)$ of each message. The merged trajectories exhibit a consistent pattern with the large-scale spreading trajectories, i.e., first rising and then falling~(see SI Sec.~IV, XI). In this sense, our Eq.~(\ref{eq:unified_equation_ultimate}) also captures the spreading mechanism of small-scale spreading messages.

As consistent spreading pattern on mainstream social media platforms such as WeChat, Weibo, Twitter, and Facebook~\cite{ugander2012structural}, we believe that Eq.~(\ref{eq:unified_equation_ultimate}) offers a unified framework for information spreading on online social networks. In this work, we analyze the retweeting behavior of a cumulative of 7.45 billion users. Although it may not be enough to claim that Eq.~(\ref{eq:unified_equation_ultimate}) is universally applicable, it suggests a widespread phenomenon in human behavior, indicating the possibility of its universality, which awaits further confirmation through analysis of more comprehensive data in the future.

\subsection{Profound influence on social media}
Lastly, we would like to investigate the pattern of information spreading given by Eq.~(\ref{eq:unified_equation_ultimate}). Specifically, three characteristics of the spreading process of a message are considered: the spreading coverage $s$ - the number of individuals who retweet the message normalized by the network size when the spreading extends globally~(also called ``outbreak size''~\cite{hu2018local}), the outbreak probability $p(s)$ - the probability that the spreading coverage is $s$~\cite{newman2010networks, hu2018local, xie2021detecting}, and the spreading speed $T$ - the average time duration it takes to propagate the information from the source user to each retweeting user~(see SI Secs.~XII, XIII for the calculations). For illustration purposes, we compare our model with the classical SIR model and the social reinforcement model through simulations. Details on simulations are provided in the Methods section.

\begin{figure*}[!htbp]
  \centering
  \includegraphics[width=16cm]{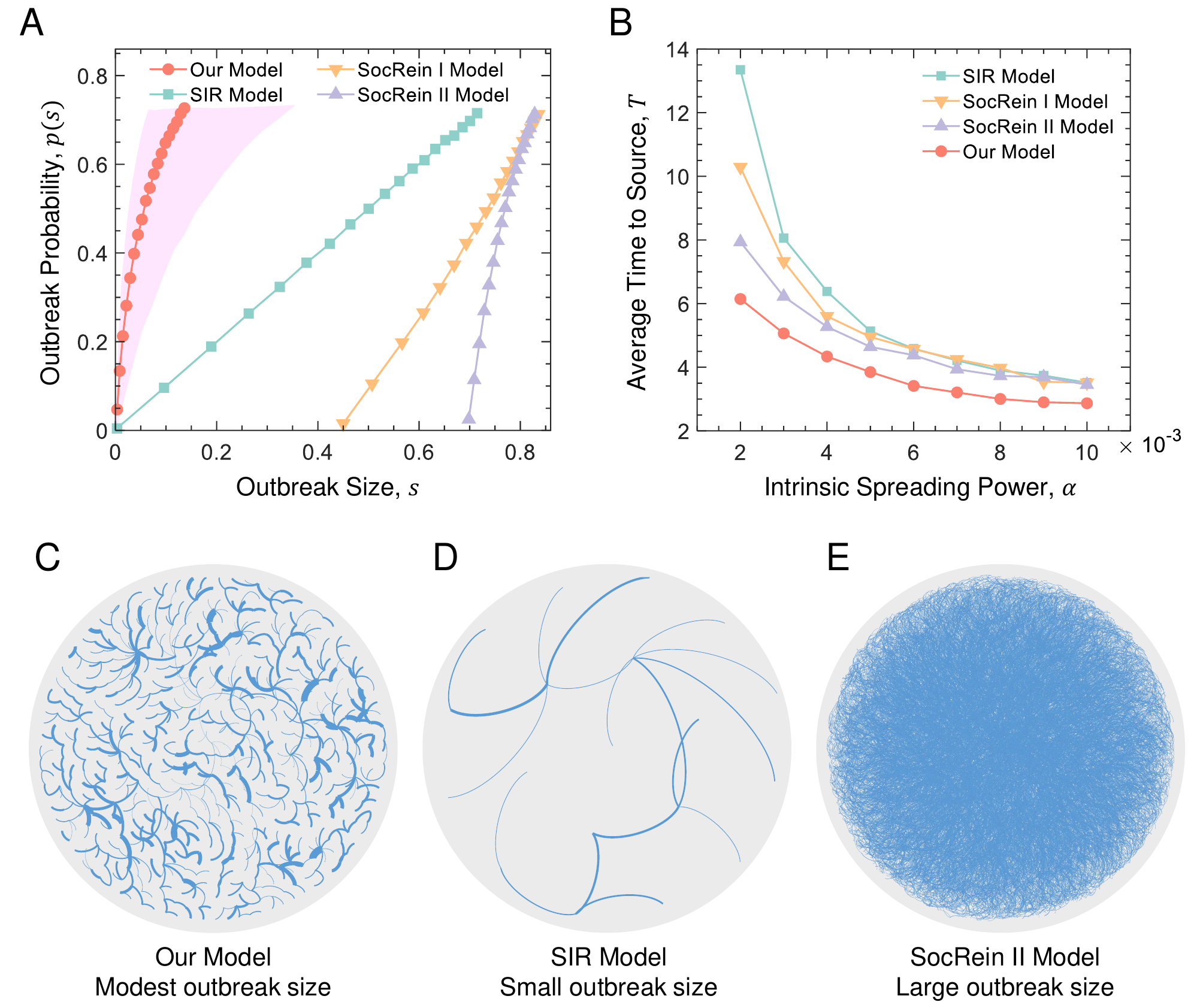}
  \caption{
  \textbf{Characteristics of information spreading process based on simulations.}
  \textbf{(A)} The outbreak probability $p(s)$ versus the spreading coverage $s$ for our model, the classical SIR model, and the social reinforcement model. ``Our model'' refers to the simulation based on our model with fixed $\omega$~($=0.95$, i.e., the averaged $\omega_i$ over 182 WeChat messages) and varying $\alpha$. The parameters of ``SocRein I model'' and ``SocRein II model'' are $b=0.01$ and $b=0.0001$, respectively. Under the SIR model~(squares), we have $s=p(s)$, which is consistent with the classical propagation theory~\cite{hu2018local,newman2010networks}. Comparing with the SIR model, the curve of $p(s)$ versus $s$ under our models~(circles) is above the diagonal line and close to the vertical axis, suggesting that $s \ll p(s)$. Under the social reinforcement model~(triangles), the curves of $p(s)$ versus $s$ are below the diagonal line and close to the vertical line $s=1$, indicating $s \gg p(s)$. This means that once a piece of information outbreaks, the spreading coverage is typically very large~(close to the entire network) under the social reinforcement model.
  \textbf{(B)} The time $T$ versus the intrinsic spreading power $\alpha$. Here, $T$ is the average time duration it takes to propagate the information from the source user to each retweeting user.
  \textbf{(C), (D), (E)} Visualization of information spreading trajectories under our model, the SIR model~(not outbreak), and the social reinforcement model II. The parameters used for the three models are: $\alpha=0.001, \gamma=0.24, \omega=0.95$, and $b=0.0001$. The thickness of a line represents the number of exposures $x$ of the user who retweeted the message along the line~(the larger the $x$, the thicker the line). It is observed that the outbreak size under our model is modest, compared to the smallest scale under the SIR model and the largest scale~(almost the entire network) under the social reinforcement model. You can see the spreading process of the three models in the supplementary movies~(see SI Sec.~XVIII). 
  }\label{Fig4}
\end{figure*}

It turns out that our model yields a small $s$ with $s \ll p(s)$~(see Fig.~\ref{Fig4}A), while $s=p(s)$ under the SIR model and a large $s$ with $s \gg p(s)$ under the social reinforcement model. Fig.~\ref{Fig4}C-E further visualize an information outbreak under our model, the SIR model, and the social reinforcement model on a network with empirical WeChat features~(see SI Sec.~XIV for details), respectively. These results imply that information spreading under our model is relatively small in outbreak size and high in outbreak probability in contrast to a large outbreak size~(almost the entire network) predicted by the social reinforcement model~\cite{centola2010spread, chen2020rumor, zheng2013spreading}. This is consistent with empirical observations and reasonably explains the contradiction raised by the social reinforcement mechanism. Furthermore, Fig.~\ref{Fig4}B demonstrates that the spreading speed is the fastest under our model, followed by the social reinforcement model, and the slowest under the SIR model. Altogether, the small outbreak size, the high outbreak probability, and the rapid information outburst facilitate a large number of information outbreaks per unit of time, which is of great significance to the diversity of information spreading on social media.

\section{Conclusion and Discussion}
In this work, we study the information propagation on social media as the propagation of human behaviors. Therein, associating the spreading probability with the gain that motivates the spreading behavior, we propose a concise and universal equation for the information spreading dynamics on social media. The equation not only fits every empirical spreading trajectory collected on large-scale social media platforms surprisingly well, but also clarifies phenomena that have failed to be explained by previous theories. Particularly, we find that the empirical retweeting probabilities show a uniquitous pattern of first rising and then falling. With the two competing mechanisms in our model, we can well explain this pattern as well as phenomena like the weak tie effect. Our model demonstrates that the highly clustered nature of real social media facilitates rapid, frequent, and relatively small-scale information bursts, enabling social media to have a high capacity and diversity for information dissemination. Our model links microscopic individual interactions to the macroscopic phenomena of social media. Due to its simplicity and adaptability, we believe that the proposed equation is likely to serve as a fundamental equation for information spreading and inspire a new wave of theoretical and applied social media research.

Despite the contributions of this study, several limitations need to be acknowledged. Firstly, during the process of information dissemination on social media, users may add new comments when forwarding a message, e.g., Twitter's quote tweet feature~\cite{cheng2018diffusion,cheng2016cascades}. However, our current model and data are insufficient to clarify how these comments influence the dissemination process. Secondly, the increasing presence of artificially manipulated bots on social media and their impact on the dissemination process is not explicitly addressed in this paper. Although we theoretically demonstrate that the proposed model can explore these two aspects~(see SI Sec.~XVI and Sec.~XVII), there is a lack of comprehensive real-world data for an in-depth discussion at present. Finally, the integration of the model presented in this paper with artificial intelligence algorithms to better predict information spreading size remains an important topic for future research~\cite{cheng2014can}.

\section{Methods}
\subsection*{Empirical data and preprocessing}
We collect empirical data from three mainstream social media platforms, namely WeChat, Weibo, and Twitter~\cite{hodas2012visibility}. The datasets are strictly de-identified so that no detailed user profiles are available. In total, 294, 8,283, and 66,064 spreading trajectories are obtained from WeChat, Weibo, and Twitter, respectively. Then, a preprocessing step called truncation is applied to the data to pick out trajectories with at least 5 data points on the retweeting probability curve $\beta(x)$, each involving no less than 100 retweeting users. The reasons to perform such data preprocessing are: (1) with too few data points on $\beta(x)$, the solution of our spreading equation Eq.~(\ref{eq:unified_equation_ultimate}) would not be unique; (2) with too few retweeting users, the computed empirical retweeting probabilities would contain too much noise. For more detailed descriptions of the empirical data, as well as sensitivity analyses on the truncation performed, please refer to SI Sec.~I and Sec.~XV A. As a result, the trajectories of 182 messages from WeChat and 100 messages from Weibo are used in the analyses of large-scale spreading messages. For Twitter, since the data is incomplete as stated in~\cite{hodas2012visibility}, we only focus on the small-scale spreading messages which are merged and show a consistent first-rises-and-then-falls pattern with the large-scale spreading messages~(see SI Sec.~XV B for the data preprocessing applied to the Twitter data).

\subsection*{Computation of the empirical retweeting probability}\label{sec:getBetaX}
The probability of users forwarding a message can be calculated based on its spreading trajectory and the associated network structure~\cite{romero2011differences, ugander2012structural, hodas2014simple}. To compute the retweeting probability defined by Eq.~(2) in the main text, how the numerator $m_i(x)$ and the denominator $r_i(x)$ are determined from empirical data are presented below. From the empirical spreading trajectory of message $i$ in WeChat, we obtain the following information for each user $j$: her total number of exposures to the message, her neighbors who posted the message, the time when her neighbors posted the message, whether she retweets the message, and if yes from which neighbor she retweets. Taking user $j$ who is exposed to message $i$ three times in total as an illustrating example, we describe how she is counted in $m_i(x)$ and $r_i(x)$. Suppose that three of her neighbors denoted as $a$, $b$, $c$, posted the message, and the time when they posted the message satisfies $t_a < t_b < t_c$. If user $j$ does not retweet the message after seeing the message three times, then she is counted in $r_i(1)$, $r_i(2)$, and $r_i(3)$. However, if user $j$ retweets from $b$, for example, then she is considered to be retweeting at the second exposure regardless of whether the retweeting occurs before or after $t_c$, meaning that she is counted in $r_i(1)$, $r_i(2)$ and $m_i(2)$, please see SI Sec.~II for more details.

\subsection*{Simulation of information spreading process}
Here we describe how the simulations of the information spreading process are performed to generate Fig.~\ref{Fig4}. The spreading process of a piece of information on a network takes the network size $N$, the adjacency matrix $\boldsymbol{A}$, and the retweeting probability $\beta(x)$~(determined by the model we use) as inputs, and outputs the spreading trajectory encoded as a matrix $\boldsymbol{T}$. Each row of $\boldsymbol{T}$ records the details of one retweeting, i.e., the ID of the retweeting user, the ID of the retweeting user's neighbor from whom she retweets, and the time of retweeting. Therefore, the dimension of $\boldsymbol{T}$ is $(s-1) \times 3$ where $s$ is the spreading coverage~(excluding the source user of the information spreading process). Put in simple terms, the simulation process is to randomly select a user in the network as the source of information spreading, and then sequentially select users that have at least one retweeting neighbor to further propagate the information with probability $\beta(x)$.
However, the concrete simulation process contains many details that need further explanation.

To elaborate on the detailed simulation process for a given piece of information, we introduce some additional notations. The set of neighbors of each user $j$ in the network is denoted as $N_j=\{k: k\neq j \text{ and } \boldsymbol{A}(j, k)=1\}$. Each user has two possible states, i.e., the S-state~(the susceptible state, i.e., the user has not retweeted the information) and the I-state~(the infected state, the user has retweeted the information). The set of S-state users with at least one I-state neighbor at a specific time point $t$ is denoted as $S_{\text{curr}}$. Assuming that the retweeting activities of these $|S_{\text{curr}}|$ users is a Poisson process, then the expected time duration from the current time point until the first retweeting occurs is $\Delta t \propto 1/|S_{\text{curr}}|$. Since our focus is on relative time duration, we set $\Delta t = 1 / |S_{\text{curr}}|$ for convenience. In addition, the following information is recorded for each user $j$ to keep track of her status at $t$: $s_j$ - the current state of the user~(S or I), $n_{Ij}$ - the number of the user's neighbors in the I-state~(i.e., the number of infected neighbors), $x_{j}$ - the number of exposures to the information before the user retweets it, $t_{Ij}$ - the time at which the user enters I-state~(i.e., the time of infection). Note that we always have $x_j \leq n_{Ij}$, and $x_j$ could be less than $n_{Ij}$ as the user may not see every retweeting of her neighbors, which is closer to the situation in practice. The detailed steps of the simulation process are:
\begin{enumerate}
  \item Initialize the spreading process: set $t=0$, $n_{Ij}=x_{j}=0$, $t_{Ij}=0$ for $j=1, 2, \ldots, N$, and $\boldsymbol{T}=[ ]$.
  \item Randomly select a user $j$ as the source node of information spreading: update her state~($s_j=I$), her neighbors' state $s_{k}=S$, her neighbors' $n_I$~($n_{Ik}=1$ for $k\in N_j$), and set $S_{\text{curr}} = N_j$.
  \item Randomly select a user $k$ from $S_{\text{curr}}$ and try to turn it into the I-state in the following way:
  \begin{enumerate}
    \item Update the current time point $t = t + 1 / |S_{\text{curr}}|$;
    \item Sort the $n_{Ik}$ infected neighbors of user $k$ by their time of infection in ascending order, select the last $m=n_{Ik} - x_k$ of them and denote the list of IDs as $v(k) = (v_1, v_2, \ldots, v_m)$;
    \item Iterate through each user $v_l$ in $v(k)$: update $x_k = x_k + 1$ and infect user $k$ with probability $\beta(x_k)$; if user $k$ is not infected by user $v_l$, continue to the next user in $v(k)$; otherwise, update user $k$'s state~($s_k=I$) and time of infection~($t_{Ik}=t$), add a row $(k, v_l, t)$ to $\boldsymbol{T}$, update user $k$'s neighbors' $n_I$ ($n_{Ir} = n_{Ir} + 1$ for $r\in N_k$) and go to step (d).
    \item Update $S_{\text{curr}} = \{r: s_r = S \text{ and } x_r < n_{Ir}\}$.
  \end{enumerate}
  \item Repeat step 3 until $S_{\text{curr}}$ is empty.
\end{enumerate}

\subsection*{Data and Code Availability}
The data and code for this study can be available upon request after publication.

\subsection*{Competing interests}
The authors declare no competing interests.

\bibliographystyle{unsrtnat}
\bibliography{Reference}



\end{document}